\documentclass[preprint,10pt]{elsarticle}
\journal{Complexity, ULB-TH/17-24 }

\usepackage{graphicx}      
\usepackage{natbib}        
\usepackage{tikz}
\usepgflibrary{arrows}
\usepackage{subfigure}
\usepackage{chemfig}
\usepackage{amssymb}
\usepackage{float}
\usepackage{amsmath}

\begin{document}

\newcommand{\E}{{\rm \bf E}}
\newcommand{\F}{{\rm \bf F}}
\newcommand{\Var}{{ \rm \bf Var}}
\newcommand{\Cov}{{\rm  \bf Cov}}
\newcommand{\Eq}{\begin{equation}}
\newcommand{\Eeq}{\end{equation}}
\newcommand{\Eqn}{\begin{eqnarray}}
\newcommand{\Eeqn}{\end{eqnarray}}

\begin{frontmatter}

\title{Deciphering noise amplification and reduction in open chemical reaction networks} 


\author{Fabrizio Pucci} 
\author{Marianne Rooman} 

\address{Department of BioModeling, BioInformatics \& BioProcesses,\\  Department of Theoretical Physics,\\ Universit\'e Libre de Bruxelles,
Roosevelt Ave. 50, 1050 Brussels, Belgium;\\  (e-mail: fapucci@ulb.ac.be, mrooman@ulb.ac.be)}
\begin{abstract}                
The impact of random fluctuations on the dynamical behavior a complex biological systems is a longstanding issue, whose understanding would shed light on the evolutionary pressure that nature imposes on the intrinsic noise levels and would allow rationally designing synthetic networks with controlled  noise.  Using the It\=o stochastic differential equation formalism, we performed both analytic and numerical analyses of several model systems containing different molecular species in contact with the environment and interacting with each other through mass-action kinetics.  These systems represent for example biomolecular oligomerization processes,  complex-breakage reactions, signaling cascades or metabolic networks.  For chemical reaction networks with zero deficiency values, which admit a detailed- or complex-balanced steady state,  all molecular species are uncorrelated. The number of molecules of each species follow a Poisson distribution and their Fano factors, which measure the intrinsic noise, are  equal to one. Systems with deficiency one have an unbalanced non-equilibrium steady state and   a non-zero S-flux, defined as the flux flowing between the complexes multiplied by an adequate stoichiometric coefficient. In this case, the noise on each species is reduced if the flux flows from the species of lowest to highest complexity, and is amplified is the flux goes in the opposite direction. These results are generalized to systems of deficiency two, which possess two independent non-vanishing S-fluxes, and we conjecture that a similar relation holds for higher deficiency systems.
\end{abstract}


\end{frontmatter}

\section{Introduction}
The identification and understanding    of the  principles that guide the modulation of intrinsic noise in biological processes is a major goal of systems biology.  Indeed, a wide range of biological phenomena such as  the biochemical reactions and the transcription and  translation machineries are of a random nature, and fluctuations play frequently a pivotal role in their dynamics \cite{Rev2,Rev2a,Rev3,Rev1}. Biological systems appear to have evolved over time to tune the noise level,  in some phenomena to reduce and tolerate the fluctuations while in others to utilize the heterogeneity to their advantage \cite{Rev2b,Rev4}. 

One of the classical examples in which the cell systems use  fluctuations to obtain a selective advantage is related to the cellular decision-making processes. Indeed, intrinsic noise can allow the diversification of the phenotype of identical cells that live in the same environmental conditions and thus facilitate the transitions between various cellular states. Multiple examples of the important role of the fluctuations in the cellular decision mechanisms in organisms of different levels of complexity - from virus and bacteria to mammalian cells - have been thoroughly analyzed in the literature (see \cite{Rev3} and references therein). 
 
In contrast, in many other biological systems, stability and robustness criteria basically require the suppression of the fluctuations and a wide series of different mechanisms are used to ensure this attenuation.  A simple and common example is the negative feedback loop in gene regulatory networks, in which the protein that is expressed from a given gene inhibits its own transcription \cite{RevUli,RevRao}. This mechanism indeed tends to suppress the noise while reducing the metabolic cost of protein production, and speeds up the rise-times of transcription units  \cite{RevFast}.

The comprehension of how the modulation of noise is achieved is very important for basically two  reasons. The first is of fundamental nature and involves  answering  open questions about why natural evolution designs  specific networks and functional mechanisms and about the role played by fluctuations  \cite{Evolution1,Evolution2}. The second reason concerns the application  to  synthetic biology with the aim of engineering and assembling biological components into synthetic devices with a controlled level of intrinsic noise \cite{Synthetic1,Synthetic2}.

Despite the many valuable advances  in the field of the last two decades, the mechanisms employed to amplify or to suppress the fluctuation levels need  to be further understood and clarified. Indeed, the huge complexity of biological systems, their dependence on a large number of variables and the system-to-system variability make the unraveling of these issues, whether using  experimental or computational approaches, a highly non-trivial task.
 
More specifically, while the noise control is relatively well understood for small and simple networks (\emph{e.g.} negative or positive feedback loops),  it is still far from clear how the fluctuations propagate through more general and complicated networks and what is the link of  the network topology and complexity with the noise buffering or amplification. Different investigations address these  issues from various perspectives, for example by characterizing the stochastic properties of the chemical reaction networks (CRNs) and  studying  the propagation of  the fluctuations  \cite{Anderson,AndersonI,AndersonII}. From a physics-oriented perspective, the authors of \cite{Udo,Polettini,Rao} analyzed the connection between the non-equilibrium thermodynamic properties of the network and the noise level. Finally, it has been shown in \cite{Cardelli} that the increase of the network complexity tends to decrease the intrinsic noise and also to reduce the effect of the extrinsic noise for some multistable model systems, whereas a dependence of the noise reduction or amplification  on the system parameters has been found in \cite{MarianneMitia}.

In this paper we expand the results presented  in \cite{PucciRooman}, where we investigated the relation between the total level of noise in various classes of CRNs with  the networks' structural characteristics. This was done by studying systems with different degrees of complexity using the It\=o stochastic differential equations formalism, and by fully exploring, both analytically and numerically, the huge parameter space of the models. Moreover, the present computational investigation goes beyond the results that we presented in \cite{PucciRooman}, since we analyzed here not only the total level of noise but also the modulation of the fluctuations for each molecular species involved in the CRNs.

\section{Chemical Reaction Networks}

In this section we review some of the basic notions of CRN theory in order to set up our conventions; for more details, see for example \cite{FeinbergI,FeinbergII,FeinbergIII}. CRNs are systems of reactions between (bio)chemical species and are characterized by triplets  $\left[\mathcal{S},\mathcal{C},\mathcal{R}\right]$. $\mathcal{S}$ represents the ensemble of all chemical species involved in the network, $\mathcal{C}$ is the set of  complexes and $\mathcal{R}$ the ensemble of  biochemical reactions. Let us consider for example the network described by:

\begin{equation}
2 A_1 \leftrightarrow A_2  \leftrightarrow A_3 + A_1.
\label{CRN1}
\end{equation}

\noindent
In this case  $\mathcal{S}= \{A_1,A_2,A_3\}$, $\mathcal{C}= \{ 2A_1, A_2, A_3+A_1\}$ and $\mathcal{R}$ are the four reactions indicated by arrows. The  reaction vectors $\subset \mathbb{R}^3$ are here equal to $[2,-1,0],[-2,1,0],[-1,1,-1],[1,-1,1]$, where the entries of the  reaction vector $i$ (with $1 \le i \le $card($\mathcal{R}$)) are equal to the stoichiometry of the molecular species $j$ (with $1 \le j \le $card($\mathcal{S}$)) in the  complexes formed or broken by  the $i$th reaction; by convention, a positive sign is  associated  to the products of a reaction and a negative sign to the reactants. For open systems, in which the molecular species are produced from or degraded to the environment, the environment is not considered as a species but  as a complex with vanishing stoichiometry coefficients. 

Three main notions have been introduced to characterize the CRNs. The first is the deficiency $\delta$ of the network defined as: 

\begin{equation}\delta = \text{card}(\mathcal{C})-  \mathcal{L} -\mathcal{X},
\end{equation}

\noindent
where $ \mathcal{L}$  is the number of linkage classes, namely the number of  connected components of the CRN, and $ \mathcal{X}$ is the dimension of the stoichiometry subspace, namely the rank of the network.  For example the network in Eq. (\ref{CRN1}) contains  three complexes, one linkage class and its  rank is equal to two, which yields $\delta=0$. 

The second notion is the reversibility of the CRN. A network is said to be $reversible$ if for each reaction connecting complex $x$ to $y$ there is an inverse reaction from $y$ to $x$. The CRN is only  $weakly$ $reversible$ if the existence of a reaction path from complex $x$ to $y$ implies the existence of a, possibly indirect, path from $y$ to $x$. 

The last notion is the complex balance. A network is complex balanced,  if, for each complex $y$, the sum of the mean reaction rates for the reactions $r \subset \mathcal{R}$ for which $y$ is a reactant complex is equal to  the sum of the mean reaction rates for $r' \subset \mathcal{R}$ for which $y$ is a product complex at the steady state $U$: 

\begin{equation}
\sum_{j\in r} \textbf{E}(a_j( \textbf{U}))  = \sum_{j \in r'} \textbf{E}(a_j( \textbf{U})). 
\label{due}
\end{equation}
Detailed balanced CRNs are a subclass of complex balanced CRNs for which this relation holds separately for each pair of forward and inverse reactions linking  two complexes. Detailed balanced steady states correspond to thermodynamic equilibrium states, whereas  the others are  non-equilibrium steady states (NESS).

In this paper we considered mass-action CRNs,  for which the the rate of a chemical reaction is proportional to the product of the concentrations (or the number of molecules) of the reactants raised to powers that are equal to their stoichiometric coefficients.
It has been shown that such CRNs are complex-balanced $if$ $and$ $only$ $if$ they are of deficiency zero and weakly reversible. This is known as the zero deficiency theorem. 

Higher deficiency CRNs correspond to systems for which $\delta$ independent conditions on the rate constants have to be satisfied in order for the system to be complex balanced. In a certain sense, $\delta$ measures the "distance" of the network from complex balancing.

\section{It\=o stochastic modeling}

To describe the time evolution of stochastic bioprocesses, we used the chemical Langevin equation (CLE), which corresponds to It\=o stochastic differential equations (SDE) \cite{Ito, Allen} driven by multidimensional Wiener processes. It\=o  SDEs are equivalent to the Fokker-Planck and master equation formalisms under some mild conditions and they are  well suited for studying  biochemical reaction networks  \cite{Allen, Moon, CLE}. Here we focused on systems containing several species, which can be produced from or degraded to the environment and interact with each other to form biomolecular complexes. These systems mimic the interaction between molecular species, such as proteins, DNA or ligands that  assemble into protein oligomers or protein-ligand and protein-DNA complexes, but also more complex interactions   between, for example,  different   cell types that coexist in the same tissue.

\begin{figure}[h!]
\begin{center}
\includegraphics[width=12cm]{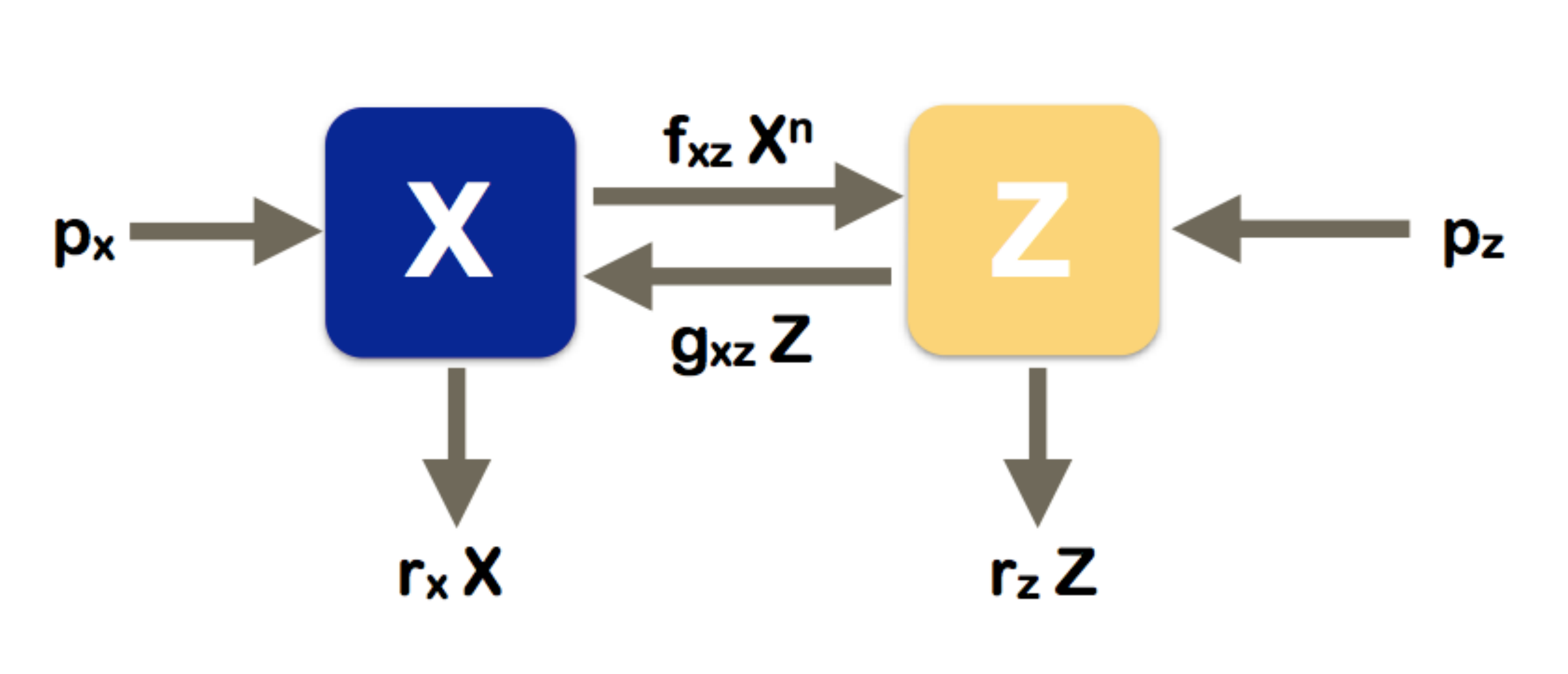}    
\caption{\textbf{Schematic picture of the reaction network  representing homooligomerization}: $n\, \text{X} \leftrightarrow \text{Z} \leftrightarrow  \varnothing \leftrightarrow  \text{X}$.}
\label{fig_CRN1}
\end{center}
\end{figure}

We start considering the open chemical reaction network depicted in Fig. \ref{fig_CRN1}, which models  for example the  process by which $n$ protein monomers $x$ assemble into a homooligomer $z$, which in turn   disassembles into $n$ monomers $x$. Both the monomers and oligomers can be produced from the environment or be degraded.

The system of It\=o SDEs that describes the dynamics of this reaction network as a function of a continuous time parameter $t \in [0, T]$ reads as: 

\Eqn
d  X(t) &=& d P_x(X,t) - d R_x(X,t) + n \left [ d  G_{xz} (X, Z, t) -  d F_{xz} (X, Z, t) \right ] \nonumber \\
d  Z(t) &=& d P_z(Z,t) - d R_z(Z,t) + d  F_{xz} (X, Z, t) - d G_{xz} (X, Z, t)  \label{A1}
\Eeqn
\noindent
where $X(t)$ and $Z(t)$ are the number of molecules of types $x$ and $z$, $d P_x$ and $d P_z$ represent the production rates for the corresponding molecular species, $d R_x$ and $d R_z$ their degradation rate, and $d F_{xz}$ and $d G_{xz}$ the interconversion terms. 

We chose the production  rates to be constant, the degradation rates to be proportional to the number of molecules, and the interconversion rates to satisfy  mass-action kinetics, thus to be  proportional to the product of the number of molecules of the reacting species raised to the powers of their stoichiometric coefficients.  This yields the following relations, each expressed as the sum of a deterministic and  a stochastic term: 

\Eqn
d P_x(X, t) & = & p_x   \, dt + \alpha_{p_x} \sqrt {p_x  }\, d W^{P_x}(t)  \nonumber \\ 
d R_x(X, t) & = & r_x  X(t)  \, dt + \alpha_{r_x} \sqrt {r_x  X(t)}\, d W^{R_x}(t)  \nonumber \\ 
d P_z(Z, t) & = & p_z   \, dt + \alpha_{p_z} \sqrt {p_z  }\, d W^{P_z}(t)  \nonumber \\ 
d R_z(Z, t) & = & r_z  Z(t)  \, dt + \alpha_{r_z} \sqrt {r_z  Z(t)}\, d W^{R_z}(t)  \nonumber \\
d F_{xz} (X, Z, t) & = & f_{xz} X( t)^{(n)}  \, dt + \alpha_{f _{xz}}\sqrt {f_{xz} X( t)^{(n)}  }\, d W^{F_{xz}}(t)  \nonumber \\ 
d G_{xz} (X, Z,  t) & = & g_{xz}  Z(t) \, dt + \alpha_{g_{xz}} \sqrt { g_{xz}  Z(t) }\, d W^{G_{xz}}(t)   
  \label{A2} 
\Eeqn

\noindent
where $X (t)^{(n)}\equiv X(t) (X(t)-1) \dots  (X(t)-n+1)$. The six $W(t)$ functions stand for independent Wiener processes,  satisfying   $W(0)=0$  with  $W(t)-W(t')$ following a $\mathcal N(0,t-t')$ normal distribution for all $(t,t')$. Note that  these processes have continuous-valued realizations and are thus appropriate when the number of molecules is large enough to be approximated as a continuous variable. In this regime we have also that $X (t)^{(n)}\simeq X (t)^n$. In what follows, we will thus always consider this approximation. The six parameters $\alpha$ that appear in front of  the stochastic terms measure the degree of stochasticity of the associated processes. For a simple birth-death process involving only one species $x$,  the  process is purely deterministic when  $\alpha_{r_x}=0=\alpha_{p_x}$,  the fluctuations follow a Poisson distribution when $\alpha_{r_x}=1=\alpha_{p_x}$. The stochasticity of the process is increased (super-Poissonian) when $\alpha_{r_x}>1$ and $\alpha_{p_x}>1$ and decreased (sub-Poissonian) when both parameters are smaller than one.

In order to solve the system  using either analytical  or  numerical techniques, we approximated the continuous-time SDEs given by Eqs (\ref{A1},\ref{A2}) by  discrete-time SDEs. Therefore, the time interval $[0,T]$ was divided into $\Xi$ equal-length intervals $0=t_0< \ldots< t_\Xi=T$, with $t_\tau=\tau \Delta t$ and $\Delta t = T/\Xi$. Using the Euler-Maruyama discretization scheme \cite{Euler}, the discrete-time SDEs read as:
\Eqn
X_{\tau +1} &=& X_{\tau} + \Delta P_x(X_{\tau}) - \Delta R_x(X_{\tau}) +n \left [ \Delta   G_{xz} (X_{\tau}, Z_{\tau}) - \Delta   F_{xz} (X_{\tau}, Z_{\tau})\right ] \nonumber \\
Z_{\tau +1} &=& Z_{\tau}  + \Delta P_z(Z_{\tau}) - \Delta R_z(Z_{\tau}) + \Delta   F_{xz} (X_{\tau}, Z_{\tau}) - \Delta   G_{xz} (X_{\tau}, Z_{\tau})   \label{A3}
\Eeqn
for all positive integers $\tau \in [0,\Xi]$, with the discretized reaction rates   given by:

\Eqn
\Delta P_x(X_{\tau}) & = & p_x   \, \Delta t + \alpha_{p_x} \sqrt {p_x  }\, \Delta W^{P_x}_\tau  \nonumber \\ 
\Delta R_x(X_{\tau}) & = & r_x X_\tau  \, \Delta t + \alpha_{r_x} \sqrt {r_x  X_\tau}\, \Delta W^{R_x}_\tau  \nonumber \\ 
\Delta P_z(Z_{\tau}) & = & p_z   \, \Delta t + \alpha_{p_z} \sqrt {p_z  }\, \Delta W^{P_z}_\tau  \nonumber \\ 
\Delta R_z(Z_{\tau}) & = & r_z  Z_\tau  \, \Delta t + \alpha_{r_r} \sqrt {r_z  Z_\tau}\, \Delta W^{R_z}_\tau  \nonumber \\ 
\Delta F_{xz} (X_{\tau}, Z_{\tau}) & = &  f_{xz} X_\tau^n  \, \Delta t + \alpha_{f_{xz}} \sqrt {f_{xz} X_\tau^{n}  }\, \Delta W^{F_{xz}}_{\tau}    \nonumber \\ 
\Delta G_{xz} (X_{\tau}, Z_{\tau}) & = &  g_{xz}  Z_\tau \, \Delta t + \alpha_{g_{xz}} \sqrt { g_{xz} Z_\tau  }\, \Delta W^{G_{xz}} _{\tau}  \label{A4} 
\Eeqn
The  independent Wiener processes satisfy  $W_\tau=W(t_\tau)$ and $\Delta W_\tau= W_{\tau+1} - W_\tau$, so that in particular $W_0=0$, $\E (\Delta W_\tau)=0$ and $\Var (\Delta W_\tau)= \Delta t $.  This system  converges towards a steady state in the long-time limit, obtained by first taking the limit $T=\Xi \Delta t \to\infty$    followed by $\Delta t \to0$. The values of the variables at the steady state will be represented without subscript, \emph{e.g.} $X_{\tau} \to X$. 

To solve analytically these SDEs, and get the mean, variances,  and covariances of the different variables at the steady state as a function of the parameters, {\it i.e.} $\E(X)$, $\Var (X)$, $\E(Z)$, $\Var (Z)$, and $\Cov(X,Z)$, we take the mean of Eqs (\ref{A3}), the mean of their squares, and the mean of the square of well-chosen combinations. The system closes for $n=1$ but not   for $n>1$, and in this case we thus need to make approximations. We used the moment closure approximation \cite{Grima} yielding for example:
\Eqn
\E(X^{n+1}) &\approx&\E(X^{n})\left ( \E(X) + n \frac {\Var(X)}{ \E(X)} \right) \nonumber\\
\E(Z X^{n}) &\approx&\E(X^{n})\left ( \E(Z) + n \frac {\Cov(X,Z)}{ \E(X)} \right )\label{closure}
 \label{closure}
\Eeqn

As usual in reaction networks, the intrinsic noise on the molecular species $x$ and $z$ is quantified through their Fano factors $\F(X)$ and $\F(Z)$,  defined as:

\Eq 
\F(X)= \frac{\Var (X)}{\E(X)} \quad , \quad \F(Z)= \frac{\Var(Z)}{\E(Z)}
\Eeq
If the $X$ and $Z$  follow a Poisson distribution, their Fano factor $\F$ is equal to one. When $\F$ is larger than one, the intrinsic noise affects more strongly the  variable concentration, and the distribution is called super-Poissonian. The distribution is  called sub-Poissonian when $\F < 1$.

The It\=o SDEs described above can be easily generalizable to model more complex processes of interest, for example those described in section 5.

\section{Homooligomerization}

The homooligomerization system that is schematically depicted in Fig. 1 is a reversible CRN for non-zero interconversion terms $f_{xz}$ and $g_{xz}$. Its deficiency is  equal to one when   $n>1$ and the two species are connected to the environment, and to zero otherwise. The It\=o SDEs that describe it are given in Eqs (\ref{A3}, \ref{A4}).   They can be solved analytically,  using the moment closure approximation of Eq. (\ref{closure}). For sake of simplicity, we assumed the equality of  all stochasticity parameters: $\alpha_{r_x}= \alpha_{r_z}=\alpha_{p_x}=\alpha_{p_z}=\alpha_{f}=\alpha_{g}=\alpha$.

In this way, we obtained  the Fano factors of $X$ and $Z$ and the covariance at the steady state expressed as a function of  $J_{xz}$, the  flux that flows between the two molecular species $x$ and $z$ multiplied by the  difference in stoichiometry coefficients between the reacting complexes:
\Eq
J_{xz}=(n-1)\left ( f_{xz} \E(X^{n})- g_{xz} \E(Z) \right )
\Eeq
In what follows, we will call this flux the S-flux. It is  zero when $n=1$, in which case the steady state is complex balanced, or when $f_{xz} \E(X^{n})= g_{xz} \E(Z) $, in which case it is detailed balanced. 
In the other cases, where the S-flux does not vanish, the system admits  a unbalanced nonequilibrium steady state.  Moreover, it is positive when the flux flows towards the complex of highest complexity, defined as  the one of highest stoichiometry. In terms of the S-flux, the Fano factors and the covariance are expressed as:


\Eqn
\F(X)&= &\alpha \left [ 1-    J_{xz} \gamma_x   \right]  \nonumber \\
\F(Z)&= &\alpha \left [ 1- J_{xz}  \gamma_z \right]  \nonumber \\
\Cov(X,Z)&= &-  \alpha \, J_{xz}  \gamma_{xz}  \label{FFC}
\Eeqn
with 
\small
\Eqn
\gamma_x &=& \frac n{D\E(X)} \left  ( r_z  (n^2 f_{xz} \frac {\E(X^{n})}{\E(X)}+  g_{xz}+ r_x+ r_z )+g_{xz}  (g_{xz}+r_x+ r_z )  \right ) \ge 0\nonumber \\
\gamma_y &=& \frac n{D\E(Z)}  n^2 f_{xz}^2 \frac {\E(X^{n})^2}{\E(X)^2} \ge  0 \nonumber \\
\gamma_{xz} &=& \frac 1D n^2 f_{xz}  \frac {\E(X^{n})}{\E(X)} ( g_{xz}  + r_z ) \ge 0
\Eeqn
and  
\Eqn
D= 2 \left (n^2 r_z f_{xz}  \frac {\E(X^{n})}{\E(X)} +
   r_x g_{xz} + r_x r_z \right ) \left (n^2  f_{xz}  \frac {\E(X^{n})}{\E(X)}+  g_{xz} + r_x+ r_z  \right )  
\nonumber
\Eeqn
\normalsize
The remaining equations  yield the mean number of molecules in terms of the parameters:
\Eqn
 (g_{xz} +r_z) \E(Y) &=& p_z+f_{xz} \E(X^{n}) \nonumber \\
n f_{xz} \E(X^{n}) + r_x \E(X)&=&  p_x+ n  g_{xz} \E(Y) \label{add}
\Eeqn

First, we observe that the Fano factors and the covariance are proportional to  the stochastic parameter $\alpha$, and thus that they vanish for deterministic systems, as expected. Furthermore, the covariance is equal to minus the S-flux multiplied by a positive coefficient. This means that when the flux flows towards the complex of highest complexity, the covariance is  negative, and when it flows towards the complex of lowest complexity, it is  positive. Finally, both Fano factors $\F(X)$ and $\F(Z)$ are equal to  $\alpha$ minus the S-flux multiplied by a positive coefficient. This means that when the S-flux is positive, the noise on both species $x$ and $z$ is reduced, whereas it is amplified when the S-flux is negative. When $\alpha=1$, the reduction or amplification is with respect to Poissonian noise. From these equations also follows:
\Eqn
\F(X)+\F(Z)=\alpha \left( 2 - \ J_{xz} \gamma \right)
\Eeqn
with the positive coefficient:
\Eqn
 \gamma=\frac { n \left  ( (r_z Z+ f_{xz} \E(X^{n}))n^2 f_{xz} \frac {\E(X^{n})}{\E(X)}+  (g_{xz}+ r_z) \E(Z) (g_{xz}+r_x+ r_z )\right )}{D\E(X)\E(Z)}
 \Eeqn
We thus recover the general relation obtained in \cite{PucciRooman} for a system of a rank 2 with deficiency $\delta=1$ and stochasticity level $\alpha=1$. We   obtained here the additional result  that, for the system described by Fig. \ref{fig_CRN1}, not only the global intrinsic noise represented by the sum of the Fano factors, but also the noise on the separate species, is amplified or reduced according to the sign of the S-flux.

To obtain the Fano factors, covariances and number of molecules as a function of the parameters only, we had to consider separately the oligomers of different degrees $n$. 

\subsection*{$n=1$:  the monomeric system}
The simplest case where each molecule $z$ is built from only one molecule $x$ represents for example molecules that undergo a conformational change or move to different cell compartments, without  interaction with other biomolecules. In this case, the deficiency is zero, the system is complex balanced and the S-flux $J_{xz}$ vanishes. No approximations are needed to solve the SDE equations.
The number of molecules is obtained  from Eq. (\ref{add}) as a function of the parameters:
\Eq
\E(X)=\frac{g_{xz}p_{x}+g_{xz}p_{z}+r_{z}p_{x}}{g_{xz}r_{x}+f_{xz}r_{z}+r_{x}r_{z}}  \quad , \quad 
\E(Z)=\frac{f_{xz}p_{x}+f_{xz}p_{z}+r_{x}p_{z}}{g_{xz}r_{x}+f_{xz}r_{z}+r_{x}r_{z}} \label{delta01}
\Eeq
and Eqs (\ref{FFC}) reduce to:
\Eq
\F(X)=\alpha \quad , \quad \F(Z)=\alpha \quad , \quad \Cov(X,Z)=0 \label{delta0}
\Eeq
In conclusion, the molecules $x$ and $z$ are uncorrelated, and both have a constant level of intrinsic noise, which is Poissonian in the case $\alpha=1$. 

We would like to emphasize that the S-flux vanishes but not the flux: for most parameter values a net flux  flows between the molecules and from and towards the environment. However, as both $x$ and $z$ molecules have the same complexity, the S-flux vanishes. The determinant of the covariance matrix is here simply equal to $\alpha^2 ( \E(X)^2+\E(Z)^2)$.

\subsection*{$n=2$: the dimerization system}
In the case in which $z$ are dimers formed of two molecules $x$, the system no longer closes and we have to use  the approximations of Eq. (\ref{closure}) to have an analytical solution. The mean number of molecules at the steady state  is then obtained as a function of the systems parameters employing Eq. (\ref{add}). This yields:
\Eqn
\E(X)&\approx &\frac{- r_x (g_{xz} +r_z) + L}{4 f_{xz} r_z}\nonumber \\
\E(Z)&\approx &\frac{4 f_{xz} r_z(p_x+2 p_z) r_z+ r_x^2 (g_{xz} +r_z)  - r_x L}{8 f_{xz} r_z^2} \label{x1}
\Eeqn
with
\Eq
L=\sqrt{r_x^2(g_{xz} +r_z)^2+8 f_{xz} r_z(p_x+2 p_z+p_x p_z)} 
\Eeq
The Fano factors and covariances are given by Eqs (\ref{FFC}) with the number of molecules given by Eqs (\ref{x1}). 

The S-flux vanishes when
\Eq
\frac{f_{xz}}{g_{xz}}\approx \frac{p_z r_x^2}{p_x^2 r_z}
\Eeq
For $f_{xz}/g_{xz}$ values smaller than this threshold, the S-flux is negative  while for larger $f_{xz}/g_{xz}$ values it is positive. Note that when the $z$ molecules are not produced or the $x$ molecules not degraded, the S-flux is always positive. In contrast, it is always negative when the $x$ molecules are not produced or the $z$ molecules not degraded. 

The Fano factors as a function of the S-flux are depicted in Figs 2a-c, for some parameter values and  stochasticity level $\alpha=1$. We would first like to stress that the numerical and analytical results are very close, which indicates that the moment closure approximation used for the analytical developments is a good approximation, at least for the tested parameter values. We observe a noise reduction for all species and parameter values when the S-flux is positive, and a noise increase for negative S-flux, as expected. We also note that decreasing the production rate of the $x$ molecules, thus  lowering the   total number of molecules in the system, amplifies this noise modulation effect.

\begin{figure}[H]
\subfigure[]{\label{fig2a}}{\includegraphics[scale=0.32]{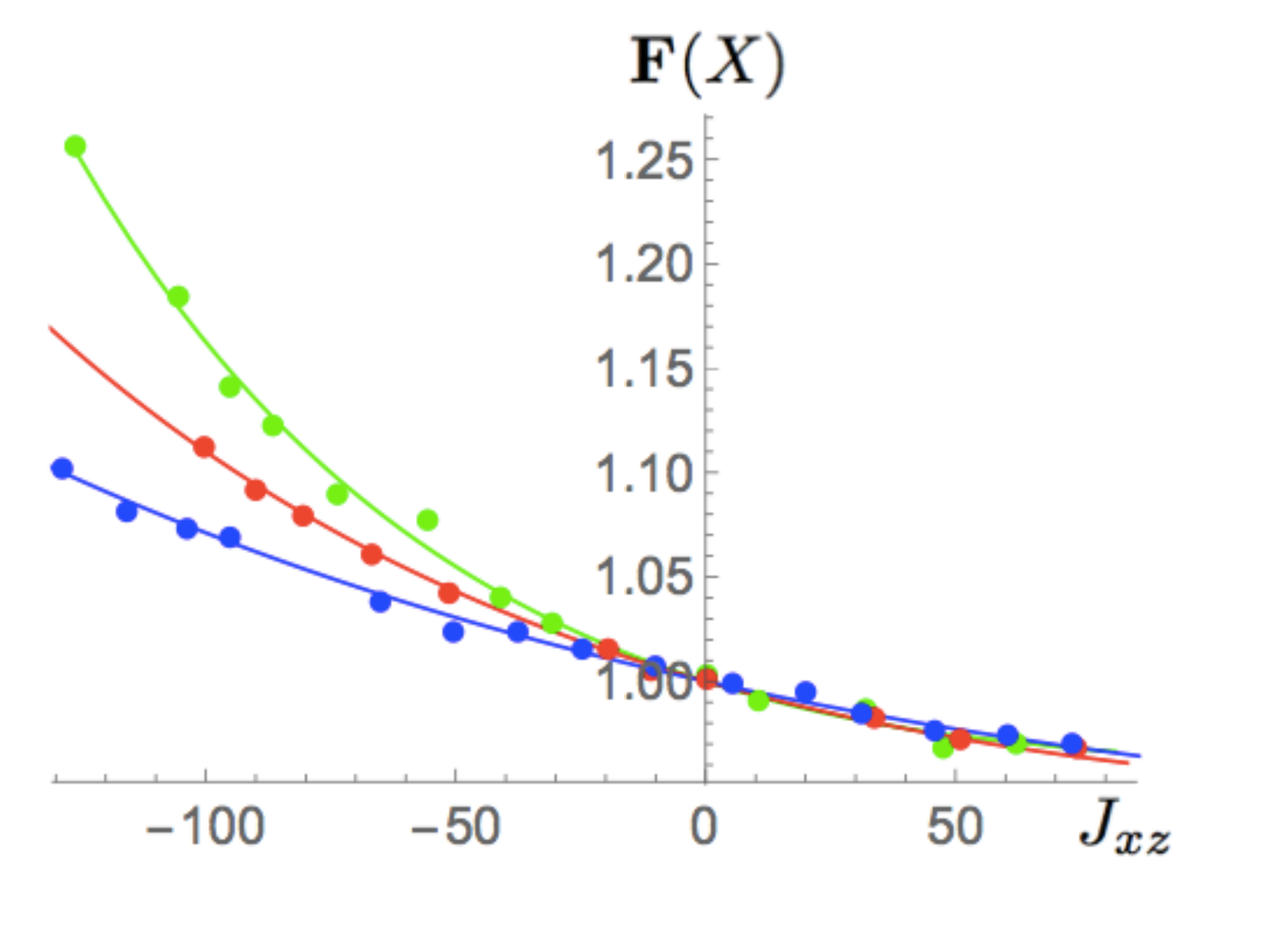}}
\subfigure[]{\label{fig2a}}{\includegraphics[scale=0.32]{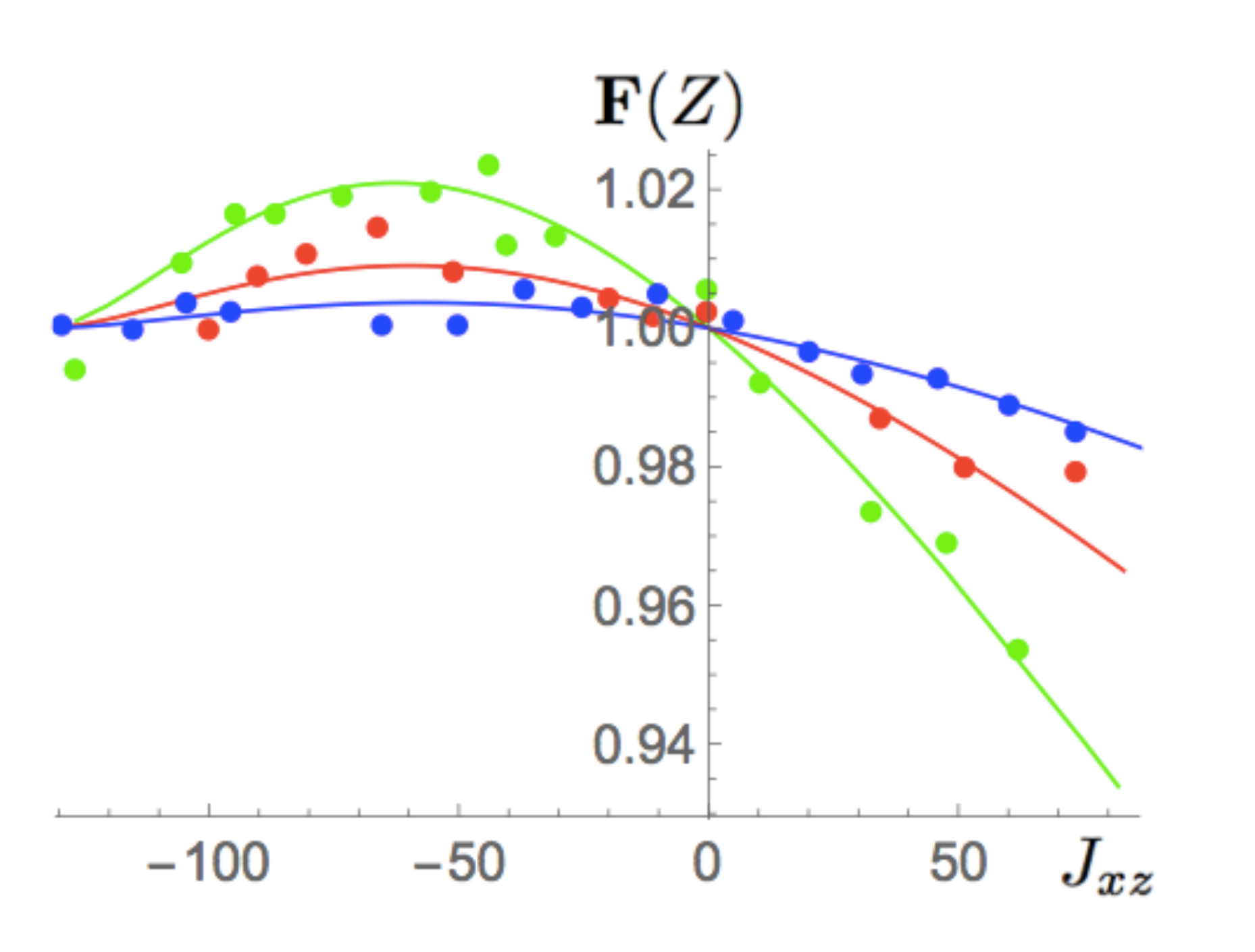}}
\subfigure[]{\label{fig2a}}{\centerline{\includegraphics[scale=0.32]{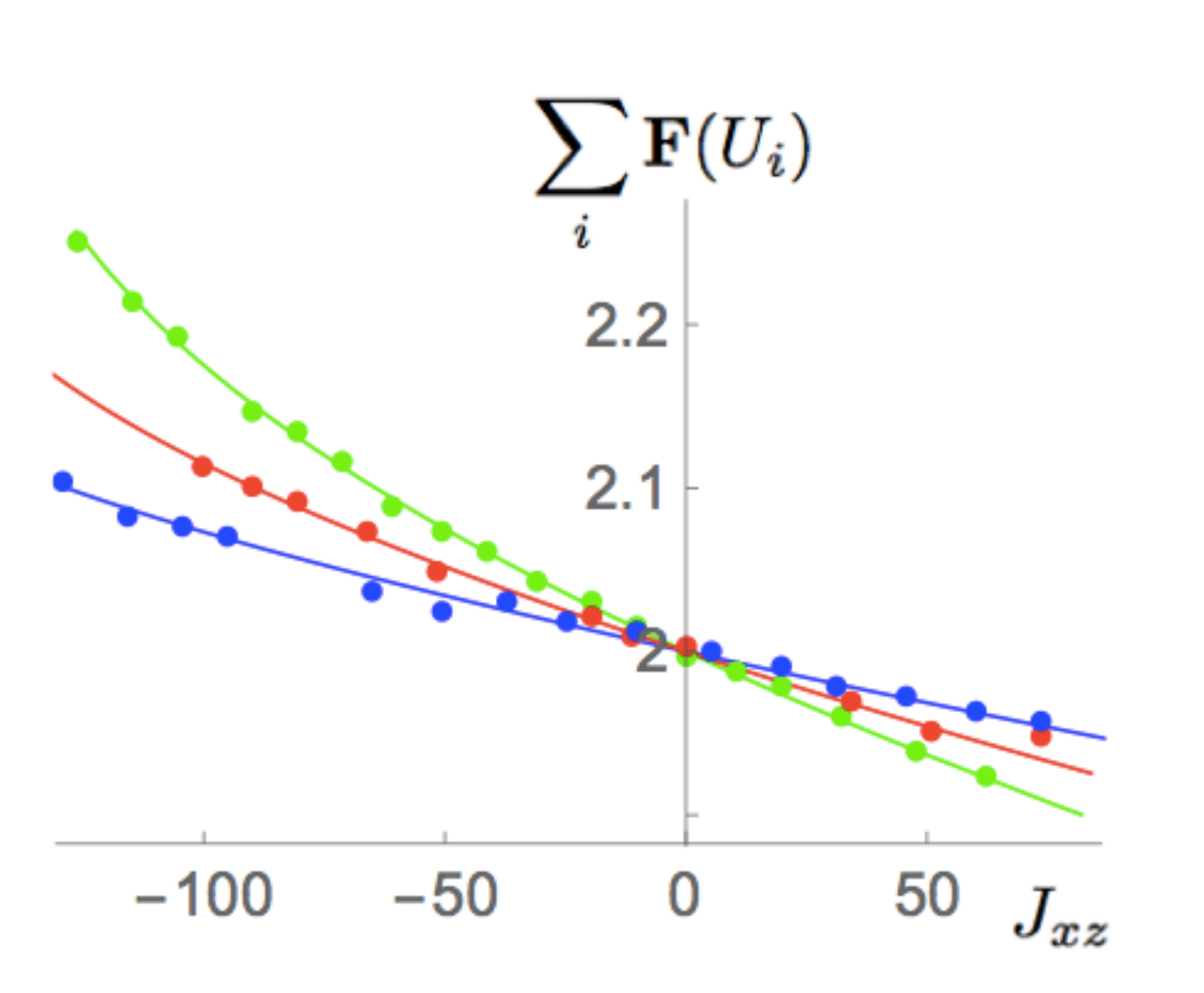}}}
\caption{\small \textbf {Stochastic behavior of homodimerization processes, as a function of the interconversion parameter $f_{xz}$}. The full lines are the analytical results of Eqs (\ref{FFC},\ref{x1}), and the points correspond to results of numerical stochastic simulations. The stochasticity parameter $\alpha=1$, the oligomerization degree $n=2$, and the parameters $p_z=200$, $r_x=r_z=0.001$ and $g_{xz}=0.002$. The parameter $p_x$ is given different values: 200 (green line), 500 (red line) and 1000 (blue line). (a)  Fano factor $\F(X)$; (b)  Fano factor $\F(Z)$; (c) Sum of Fano factors $\F(X)+\F(Z)$. }
\end{figure}


\subsection*{$n=3$ and $n=4$: the trimerization and tetramerization systems}
When $z$ are trimers or tetramers of $x$ molecules, we can use the same procedure as in the $n=2$ case, and solve the mean number of molecules at the steady state from Eq. (\ref{add}). The analytical results are given in appendix A1.

The  Fano factors of the species involved in the tetramerization process ($n=4$) are plotted as a function of the S-flux in Figs 3 a-c, and are compared with those of the dimerization ($n=2$) and the monomeric interconversion ($n=1$). Clearly, for the same parameter values, the amplification and reduction of the intrinsic noise is increased for higher oligomerization degrees. Note  the different behaviors of the Fano factors of the monomers and oligomers. When the S-flux is negative, the noise amplification on the oligomers appears limited, in contrast to the noise on the monomers which continues to grow for decreasing flux values. Instead, when the S-flux is positive, the fluctuations  of the oligomers seems to be suppressed more strongly than those  of the monomers.

\begin{figure}[H]
\subfigure[]{\label{fig2a}}{\includegraphics[scale=0.3]{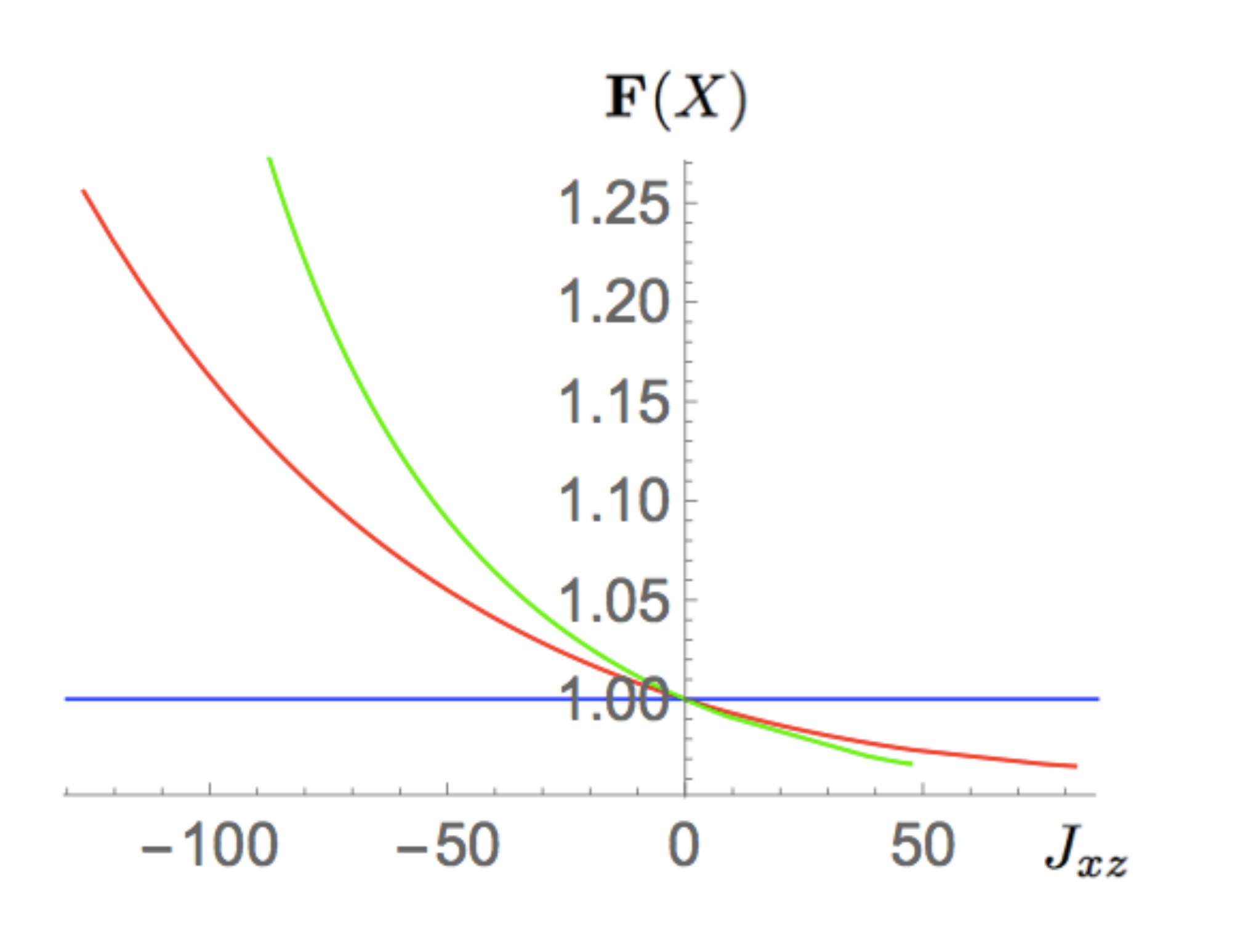}}
\subfigure[]{\label{fig2a}}{\includegraphics[scale=0.3]{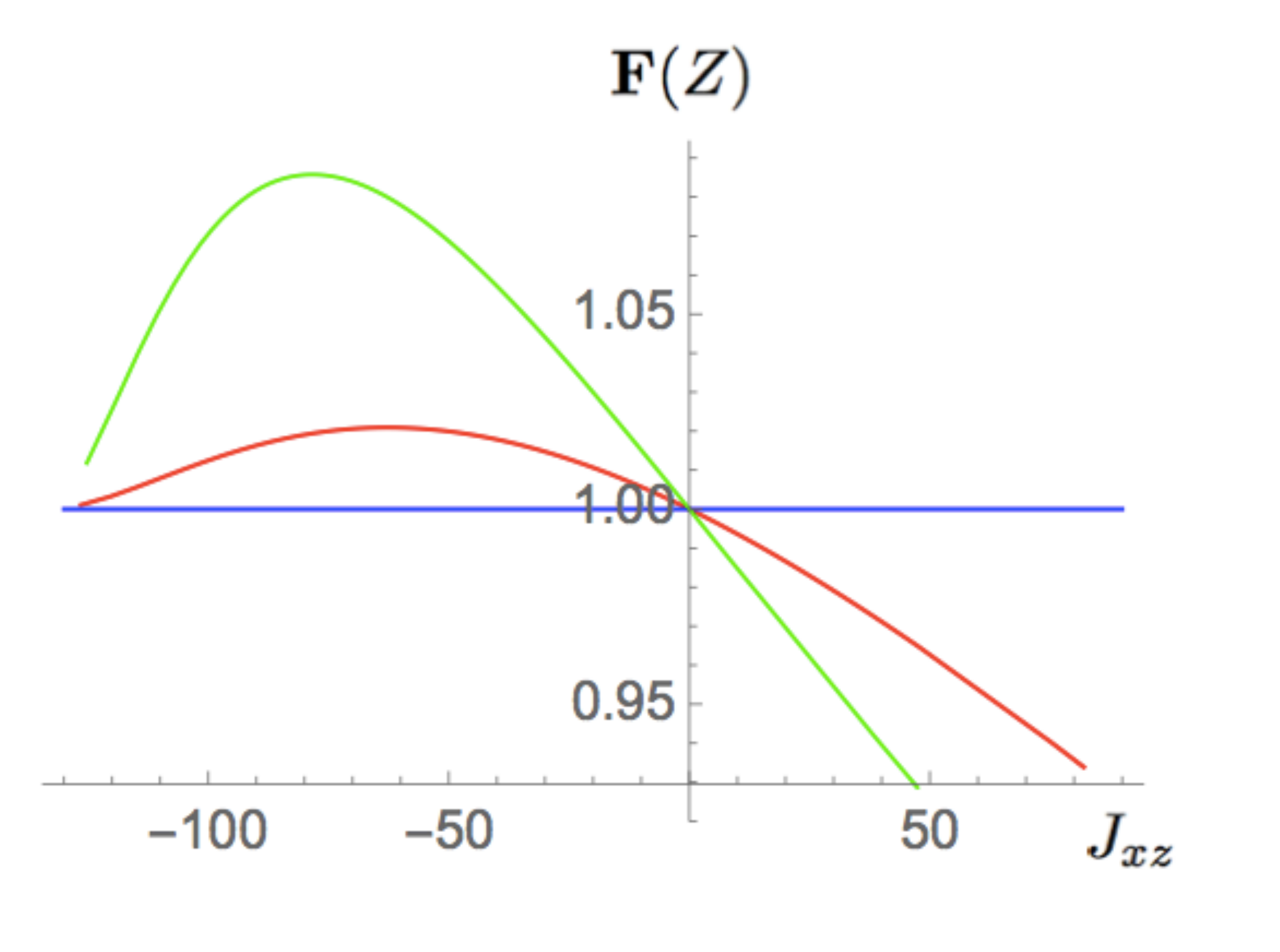}}
\subfigure[]{\label{fig2a}}{\centerline{\includegraphics[scale=0.3]{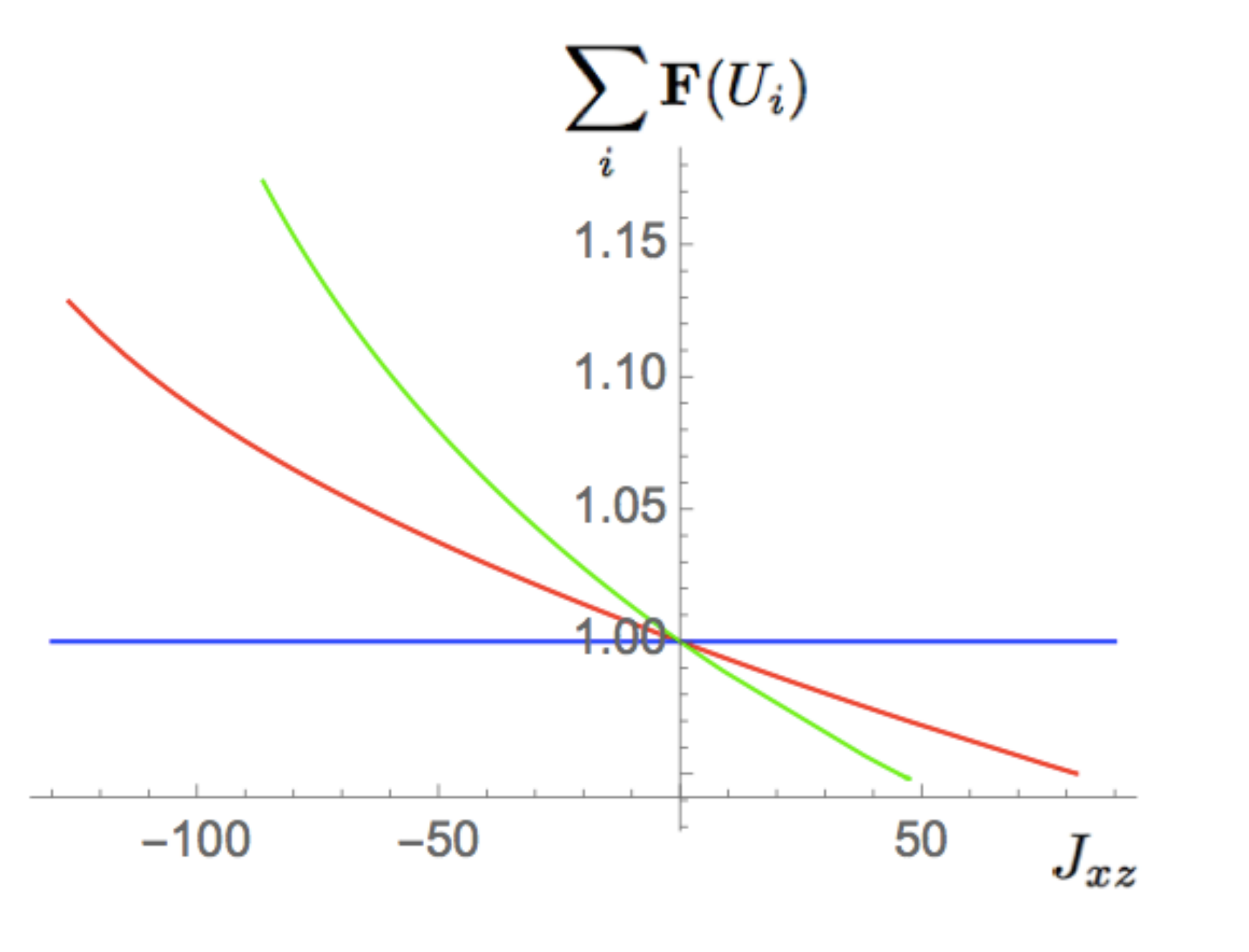}}}
\caption{\small \textbf {Stochastic behavior of homooligomerization processes, as a function of the interconversion parameter $f_{xz}$}. The full lines are the analytical results of Eqs (\ref{FFC},\ref{x1},\ref{n4}). The stochasticity parameter $\alpha=1$, and the parameters $p_x=p_z=200$, $r_x=r_z=0.001$ and $g_{xz}=0.002$. The oligomerization degree is given different values: $n=1$ (blue line), $n=2$ (red line) and $n=4$ (green line). (a)  Fano factor $\F(X)$; (b)  Fano factor $\F(Z)$; (c) Sum of Fano factors $\F(X)+\F(Z)$. }
\end{figure}

\section{Oligomerization reactions with intermediate steps}

\begin{figure}[H]
\begin{center}
\includegraphics[width=13cm]{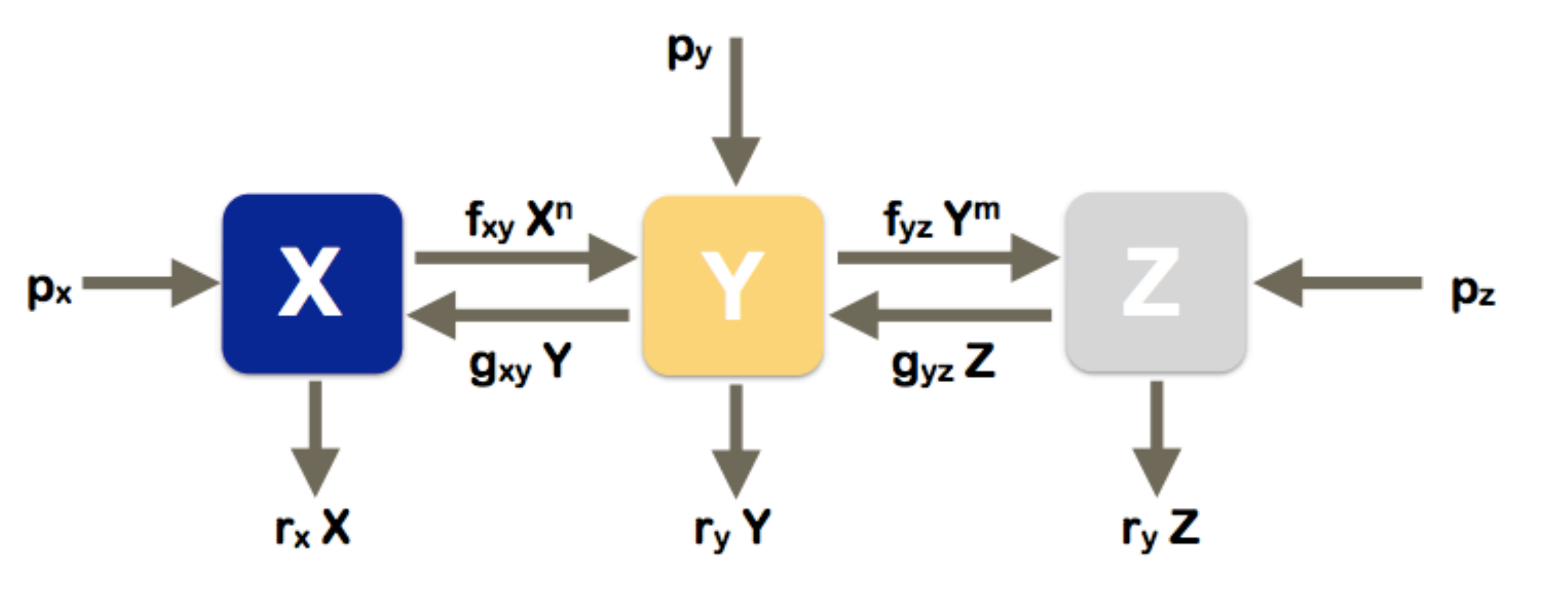}    
\caption{\bf{Schematic picture of the reaction network  representing homooligomerization with an intermediate state}: $n\, \text{X} \leftrightarrow \text{Y}$, $m\, \text{Y} \leftrightarrow \text{Z}$, $\text{Z} \leftrightarrow  \varnothing,  \text{X} \leftrightarrow  \varnothing,   \text{Y} \leftrightarrow  \varnothing$.}
\label{fig_CRN2}
\end{center}
\end{figure}

We now turn to the more complex systems schematically depicted in Fig. \ref{fig_CRN2}. They  describe a wide range of biological systems such as monomeric proteins that tetramerize through an intermediate step of dimerization \cite{IntermediateI} or that undergo amyloid formation through oligomeric intermediates \cite{IntermediateII}.  

Such systems are  reversible CRNs for non-zero values of the interconversion parameters $f_{xy}$, $f_{yz}$, $g_{xy}$ and $g_{yz}$,  with  deficiency values up to $\delta=2$. In particular,  when all the species are connected to the environment, we have $\delta=2$ for $n>1$ and $m>1$, $\delta=1$ when either ($n>1,m=1$) or ($n=1,m>1$), and $\delta=0$ for $n=1=m$. These CRNs admit a non-equilibrium steady state which is complex or detailed balanced  if $\delta=0$. 

To model these systems, we used the same formalism as in the previous section, namely discrete-time  It\=o  SDEs with an Euler-Maruyama discretization scheme \cite{Euler}. The system of non-linear coupled SDEs reads as:
\Eqn
X_{\tau +1} &=& X_{\tau} + \Delta P_x(X_{\tau}) - \Delta R_x(X_{\tau}) +n \left [ \Delta   G_{xy} (Y_{\tau}) - \Delta   F_{xy} (X_{\tau})\right ] \nonumber \\
Y_{\tau +1} &=& Y_{\tau} + \Delta P_y(Y_{\tau}) - \Delta R_y(Y_{\tau}) + \left [ \Delta   F_{xy} (X_{\tau}) -\Delta   G_{xy} (Y_{\tau}) \right ]+ m\left [ \Delta   G_{yz} (Z_{\tau}) -  \Delta   F_{yz} (Y_{\tau})\right ] \nonumber \\
Z_{\tau +1} &=& Z_{\tau}  + \Delta P_z(Z_{\tau}) - \Delta R_z(Z_{\tau}) + \Delta   F_{yz} (Y_{\tau}) - \Delta   G_{yz} (Z_{\tau})   \label{A5}
\Eeqn
 for all positive integers $\tau \in [0,\Xi]$. The discretized reaction rates  are given by:

\Eqn
\Delta P_x(X_{\tau}) & = & p_x   \, \Delta t + \alpha_{p_x} \sqrt {p_x  }\, \Delta W^{P_x}_\tau  \nonumber \\ 
\Delta R_x(X_{\tau}) & = & r_x X_\tau  \, \Delta t + \alpha_{r_x} \sqrt {r_x  X_\tau}\, \Delta W^{R_x}_\tau  \nonumber \\ 
\Delta P_y(Z_{\tau}) & = & p_y   \, \Delta t + \alpha_{p_y} \sqrt {p_y  }\, \Delta W^{P_y}_\tau  \nonumber \\ 
\Delta R_y(Z_{\tau}) & = & r_y  Y_\tau  \, \Delta t + \alpha_{r_y} \sqrt {r_y  Y_\tau}\, \Delta W^{R_y}_\tau  \nonumber \\ 
\Delta P_z(Z_{\tau}) & = & p_z   \, \Delta t + \alpha_{p_z} \sqrt {p_z  }\, \Delta W^{P_z}_\tau  \nonumber \\ 
\Delta R_z(Z_{\tau}) & = & r_z  Z_\tau  \, \Delta t + \alpha_{r_z} \sqrt {r_z  Z_\tau}\, \Delta W^{R_z}_\tau  \nonumber \\ 
\Delta F_{xy} (X_{\tau}) & = &  f_{xy} X_\tau^n  \, \Delta t + \alpha_{f_{xy}} \sqrt {f_{xy} X_\tau^{n}  }\, \Delta W^{F_{xy}}_{\tau}    \nonumber \\ 
\Delta G_{xy} (Y_{\tau}) & = &  g_{xy}  Y_\tau \, \Delta t + \alpha_{g_{xy}} \sqrt { g_{xy} Y_\tau  }\, \Delta W^{G_{xy}} _{\tau}  \nonumber \\ 
\Delta F_{yz} (Y_{\tau}) & = &  f_{yz} Y_\tau^m  \, \Delta t + \alpha_{f_{yz}} \sqrt {f_{yz} Y_\tau^{m}  }\, \Delta W^{F_{yz}}_{\tau}    \nonumber \\ 
\Delta G_{yz} (Z_{\tau}) & = &  g_{yz}  Z_\tau \, \Delta t + \alpha_{g_{yz}} \sqrt { g_{yz} Z_\tau  }\, \Delta W^{G_{yz}} _{\tau}  \label{A6} 
\Eeqn
where the  ten  Wiener processes are independent.  

These equations can be solved analytically,  using the moment closure approximation of Eq. (\ref{closure}). For simplicity, we again assumed the equality of  all stochasticity parameters: $\alpha_{r_x}= \alpha_{r_y}= \alpha_{r_z}=\alpha_{p_x}=\alpha_{p_y}=\alpha_{p_z}=\alpha_{f_{xy}}=\alpha_{g_{xy}}=\alpha_{f_{yz}}=\alpha_{g_{yz}}=\alpha$.
There are two S-fluxes in this CRN, which are independent and in general non-zero when $\delta=2$:
\Eqn
J_{xy}&=&(n-1)\left ( f_{xy} \E(X^{n})- g_{xy} \E(Y) \right ) \nonumber \\ 
J_{yz}&=&(m-1)\left ( f_{yz} \E(Y^{m})- g_{yz} \E(Z) \right )
\Eeqn
We obtained  the Fano factors of $X$, $Y$ and $Z$ and the covariances $\Cov(X,Y)$, $\Cov(X,Z)$ and $\Cov(Y,Z)$ at the steady state expressed as a function of these two S-fluxes:
\Eqn
\F(X)&= &\alpha \left[ 1-    J_{xy} \gamma^{xy}_x  -    J_{yz} \gamma^{yz}_x   \right]  \nonumber \\
\F(Y)&= &\alpha \left[ 1-    J_{xy} \gamma^{xy}_y  -    J_{yz} \gamma^{yz}_y     \right]  \nonumber \\
\F(Z)&= &\alpha \left[ 1-    J_{xy} \gamma^{xy}_z  -    J_{yz} \gamma^{yz}_z   \right]  \nonumber \\
\Cov(X,Y)&= &-  \alpha \, \left[  J_{xy} \gamma^{xy}_{xy}  +  J_{yz} \gamma^{yz}_{xy}   \right] \nonumber \\
\Cov(X,Z)&= &-  \alpha \, \left[  J_{xy} \gamma^{xy}_{xz}  +  J_{yz} \gamma^{yz}_{xz}   \right] \nonumber \\
\Cov(Y,Z)&= &-  \alpha \, \left[  J_{xy} \gamma^{xy}_{yz}  +  J_{yz} \gamma^{yz}_{yz}   \right]  \label{FFC1}
\Eeqn
with all $\gamma$'s positive functions of the parameters and the mean values $\E(X)$, $\E(Y)$ and $\E(Z)$. A corollary result is that the sum of the Fano factors over all species is equal to the rank $ \mathcal{X}$ of the system minus a linear combination of the S-fluxes with positive coefficients:
\Eq
\F(X)+\F(Y)+\F(Z)=\alpha \left [ \mathcal{X} -    J_{xy} \gamma^{xy}  -    J_{yz} \gamma^{yz}   \right]  
\label{sumfano}
\Eeq
The values of the positive coefficient $\gamma^{xy}$ and $\gamma^{yz}$ are explicitly given in appendix A2. We thus recover the result obtained in \cite{PucciRooman} and generalize it to the Fano factors of each species taken individually.

To get also $\E(X)$, $\E(Y)$ and $\E(Z)$  in terms of the parameters of the system, we need to solve the following relations:
\Eqn
p_x+n  \,p_y+n \,m \, p_z&=&r_x \E(X)+ n  \,r_y \E(Y)+n  \,m  \,r_z \E(Z) \nonumber \\
 p_x+n  \,g_{xy} \E(Y) &=& r_x \E(X)+ n  \,f_{xy} \E(X^n) \nonumber \\
p_z+n \, f_{yz} \E(Y^m) &=& r_z \E(Z)+ m \, g_{xy} \E(Z)\label{add1}
\Eeqn
which correspond to the mean of Eqs (\ref{A5}) at the steady state. For solving these equations, we have to  specify the  values of $m$ and $n$. 

\begin{figure}[H]
\begin{center}
\includegraphics[width=12.5cm]{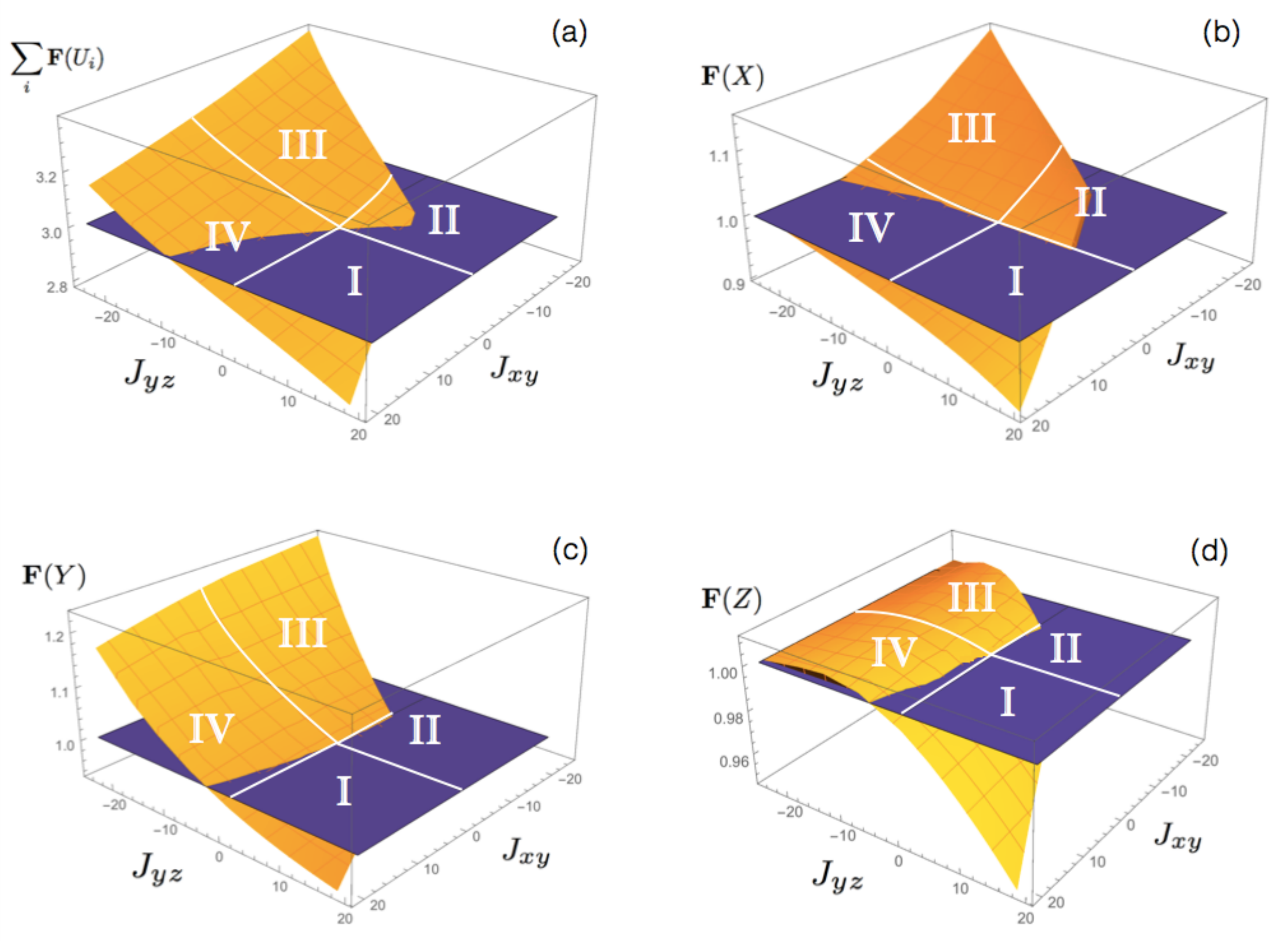}    
\caption{\textbf {Stochastic behavior of homotetramerization processes with an intermediate dimerization step, as a function of the interconversion parameters $f_{xy}$ and $f_{yz}$}.  The oligomerization degrees $n=2=m$, the stochasticity parameter $\alpha=1$, and the parameters $p_x=p_y=p_z=50$ and $r_x=r_y=r_z=g_{xy}=g_{yz}=0.001$. Fano factors: (b) $\F(X)$, (c) $\F(Y)$, (d) $\F(Z)$), and (a)  $\sum_i \F(U_i)=\F(X)+\F(Y)+\F(Z)$ as a function of the S-fluxes $J_{xy}$ and $J_{yz}$. The orange surfaces correspond to the Fano factors and the blue ones to the constant $z=1$ plane in (b-d) and $z=3$ in (a).}
\label{Intermediate}
\end{center}
\end{figure}

As an example, we analyzed the results in the case  $n=2=m$. More specifically, we plotted the values of the Fano factors of $X$, $Y$ and $Z$ as well as their sum as a function of the two fluxes $J_{xy}$ and $J_{yz}$ for $\alpha =1$, for fixed values of $p_x=p_y=p_z$,   $r_x=r_y=r_z$ and $g_{xy}=g_{yz}$, leaving the two interconversion terms $f_{xy}$ and $f_{yz}$ as free parameters. As seen in Figs \ref{Intermediate},  the number of molecules of each species follow a sub-Poissonian distribution ($\F(U_i) <1$) in the first quadrant (when both fluxes are positive) in which case the noise is reduced,  while in the third quadrant a super-Poisonnian distribution is observed with an increase of the noise level ($\F(U_i) >1$).  The domains of existence of the solutions are investigated in Appendix B.

\section{Discussion} 

In this paper, we gained important insights into the relation between the complexity of a system and its intrinsic noise, even though the picture is not yet complete. On the basis of the results obtained for the open model CRNs depicted in Figs \ref{fig_CRN1} and \ref{fig_CRN2}, we propose the following conclusions and tentative generalizations: 

\begin{itemize}
\item The modulation of noise in a mass-action CRN is related to its deficiency, and can be expressed as a function of the S-fluxes. 
\item For $\delta$=0 we have: 
\Eq\F(U_i) = \alpha
\Eeq
where $U_i$ is the number of molecules of species $i$. When $\alpha$=1, the number of molecules of each species thus follows a Poisson distribution \cite{Anderson}. Note that this is only true for open systems. For closed systems, we showed in \cite{mathmod} the weaker result $ \sum_i \F(U_i) =  \, \chi $ since in this case, the number of molecules follow a multinomial distribution  constrained by the conservation of the total number of molecules \cite{Anderson}.               
\item For $\delta$=1, there is  one independent S-flux $J_{1}$, and the Fano factors of the different species $i$ are expressed as:
\Eq
\F(U_i)=  \alpha\left [1 - J_{1} \gamma_{i}^1 \right ]
\Eeq                  
where all  the $\gamma$ coefficients are positive functions of the parameters. The noise is thus amplified when the S-flux is negative, which means that the flux flows towards the species of smallest complexity. The noise is reduced when the S-flux is positive, thus when  the flux flows towards the species of highest complexity. Note that in the case of several dependent fluxes, the positivity of the $\gamma$ coefficient is not ensured; this case will be considered in a forthcoming publication.
\item For $\delta$=2, there are  two independent S-fluxes $J_{1}$ and $J_{2}$, and the Fano factors satisfy the relations: 
\Eq
\F(U_i)=  \alpha\left [1 - J_{1} \gamma_{i}^1- J_{2} \gamma_{i}^2 \right ]
\Eeq                  
with positive $ \gamma$ values. When the two S-fluxes are positive, and thus the two fluxes flow towards the highest complexity species,  we have noise reduction on all species. When the two S-fluxes are negative and the fluxes flow towards  lowest complexity, we observe noise amplification on all species. When one S-flux is positive and the other negative, the result depends on the relative value of the associated $\gamma$ values. 
\item We argue that these trends remain valid for any value of  $\delta$, and that we have:
\Eq
\F(U_i)=  \alpha\left [1 - \sum_jJ_{j} \gamma_{i}^j \right ]
\Eeq
where the sum is over the internal  S-fluxes of the CRN, and all $\gamma$ coefficients are positive functions of the parameters.

\end{itemize}

In addition to rigorously demonstrating this conjecture for complex CRNs with generic $\delta$ values, we would like to investigate two other points. The first  is the extension of our study to systems with generalized kinetic schemes. Indeed, mass-action kinetics is only valid in the case of elementary processes occurring in homogenous solutions.  \emph{In vivo} biomolecular reactions are usually not elementary and are affected by macromolecular crowding. Their description thus requires a modification of the rate law \cite{NonMassActionI,NonMassActionII}. 

The second point is related to the modeling of noise in systems described using model reduction techniques. Indeed, systems such as metabolic or signaling networks are far too complex to be mathematically described with full details, which would require a huge number of parameters. To cope with this issue,  different reduction techniques have been introduced \cite{ModelReductionI,ModelReductionII}, such as the quasi steady-state approximation (QSSA) in which the fast variables are separated from the slow variables, and only the latter are considered as dynamical. Another reduction technique is  the variable lumping method in which the vector of the reactants is dimensionally reduced to a vector of pseudospecies, in such a way that the kinetic equations are easier to solve, and fewer parameters need to be determined. However,  it is  not trivial to deal with the fluctuations in such reduced models.  Indeed, while in elementary processes the fluctuations can be considered to follow Poisson-type distributions with all stochasticity parameters $\alpha=1$, for non-elementary reduced variables this cannot be assumed \emph{a priori}.

\section*{Acknowledgments}
We thank Mitia Duerinckx for useful discussions. FP is Postdoctoral researcher and MR Research Director at the Belgian Fund for Scientific Research (FNRS). We declare that there is no conflict of interest regarding the publication of this manuscript.

\newpage 

 

\section*{Appendix A  : Analytical Results}
\subsection*{A1. Oligomerization without  intermediate step}

For the CRN depicted in Fig. \ref{fig_CRN1}, with oligomerization degree $n$,  the  number of molecules of type $x$ and $z$ can be obtained as a function of the parameters by solving Eq. (\ref{add}). For $n=1$ and $n=2$, the solution is given in Eqs (\ref{delta01},\ref{x1}). For $n=3$, we get:
\small
\Eqn
\E(X)&=&\frac 1D \left (-2 f_{xz} r_x r_z (g_{xz} + r_z) + 
   2^{1/3}  \left (9 f_{xz}^2  r_z^2  \left( g_{xz} (p_x+3\, p_z)+ p_x r_z\right ) +L\right ) ^{2/3}\right) \nonumber \\
   \E(Y) &=& \frac{p_z+f_{xz} \E(X)^3}{ g_{xz} +r_z} \label{n3}
\Eeqn        
\normalsize
with 
\small
\Eqn
L&=&\sqrt{
      f_{xz}^3 r_z^3 \left (4 r_x^3 (g_{xz} + r_z)^3 + 
         81 f_{xz} r_z (g_{xz}(p_x + 3 p_z) + p_x r_z)^2\right )}\nonumber \\
 D&=&  2^{2/3}\, 3\,
    f_{xz} r_z  \left (9 f_{xz}^2  r_z^2  \left( g_{xz} (p_x+3\, p_z)+ p_x r_z\right ) +L\right ) ^{1/3}
    \Eeqn
\normalsize

In the case $n=4$, we have:
\small
\Eqn
\E(X)&=& \frac 1{2 \,\,2^{5/6} 3^{1/3}}\left(  \sqrt{\frac{K}{D} - \frac D {f_{xz} r_z}+\frac{6 \sqrt{2}\, r_x (g_{xz} + r_z)\sqrt{D}}{\sqrt{f_{xz}r_z}\sqrt{D^2-f_{xz} r_z K}}} - \sqrt{ \frac{D^2- f_{xz} r_z K}{f_{xz} r_z D}} \right)\nonumber \\
\E(Y) &=& \frac{p_z+f_{xz} \E(X)^4}{ g_{xz} +r_z} \label{n4}
\Eeqn        
\normalsize

with 
\small
\Eqn
L&=& \sqrt{3} \sqrt{
  f_{xz}^2  r_z^2 (27 \,r_x^4 (g_{xz} +  r_z)^4 + 1024 \,g_{xz}^4  f_{xz} r_z(g_{xz}(p_x + 4 p_z) + p_x r_z)^3)} \nonumber \\
 D&=& \left( 9 \, f_{xz} r_x^2  r_z (g_{xz} +  r_z)^2 +L \right ) ^{1/3} \nonumber \\
 K&=&8\,\,6^{1/3} (g_{xz}(p_x + 4 p_z )+ p_x r_z)
\Eeqn
\normalsize

\subsection*{A2. Oligomerization with  intermediate step}

For the CRNs depicted in Fig. \ref{fig_CRN2}, with oligomerization degrees $n$ and $m$, we obtained the Fano factors as a function of the S-fluxes multiplied by positive functions of the parameters (Eqs (\ref{FFC1},\ref{sumfano})). In particular, the positive coefficients $\gamma_{xy}$ and $\gamma_{xz}$ that appear in the sum of Fano factors, Eq. (\ref{sumfano}) are equal to: 

\noindent
$\\ \gamma_{xy}=$
\begin{figure}[H]
\begin{center}
\includegraphics[width=13cm]{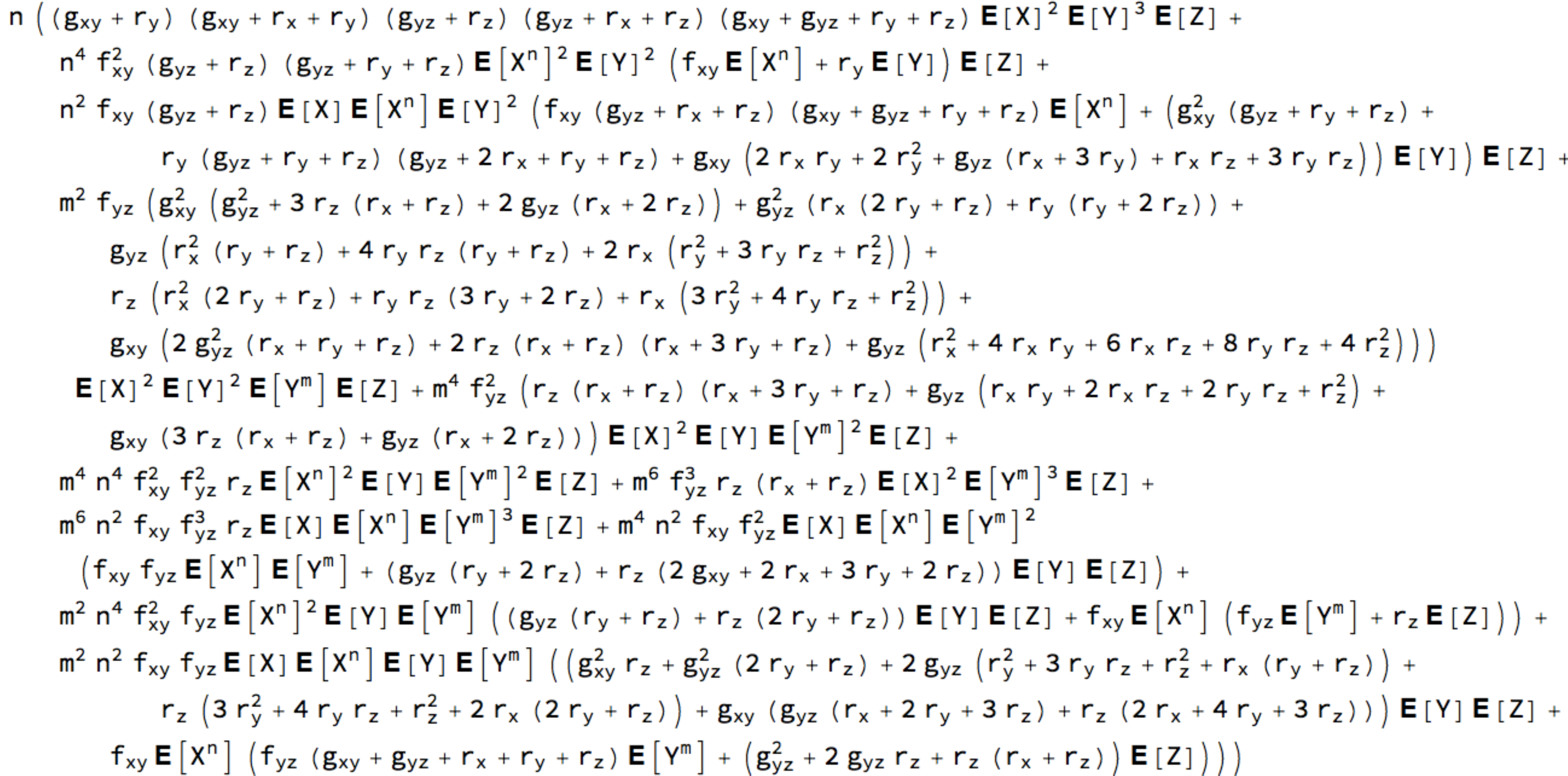}    
\label{gammaxy}
\end{center}
\end{figure}

\noindent
$\\ \gamma_{yz}=$
\begin{figure}[H]
\begin{center}
\includegraphics[width=13cm]{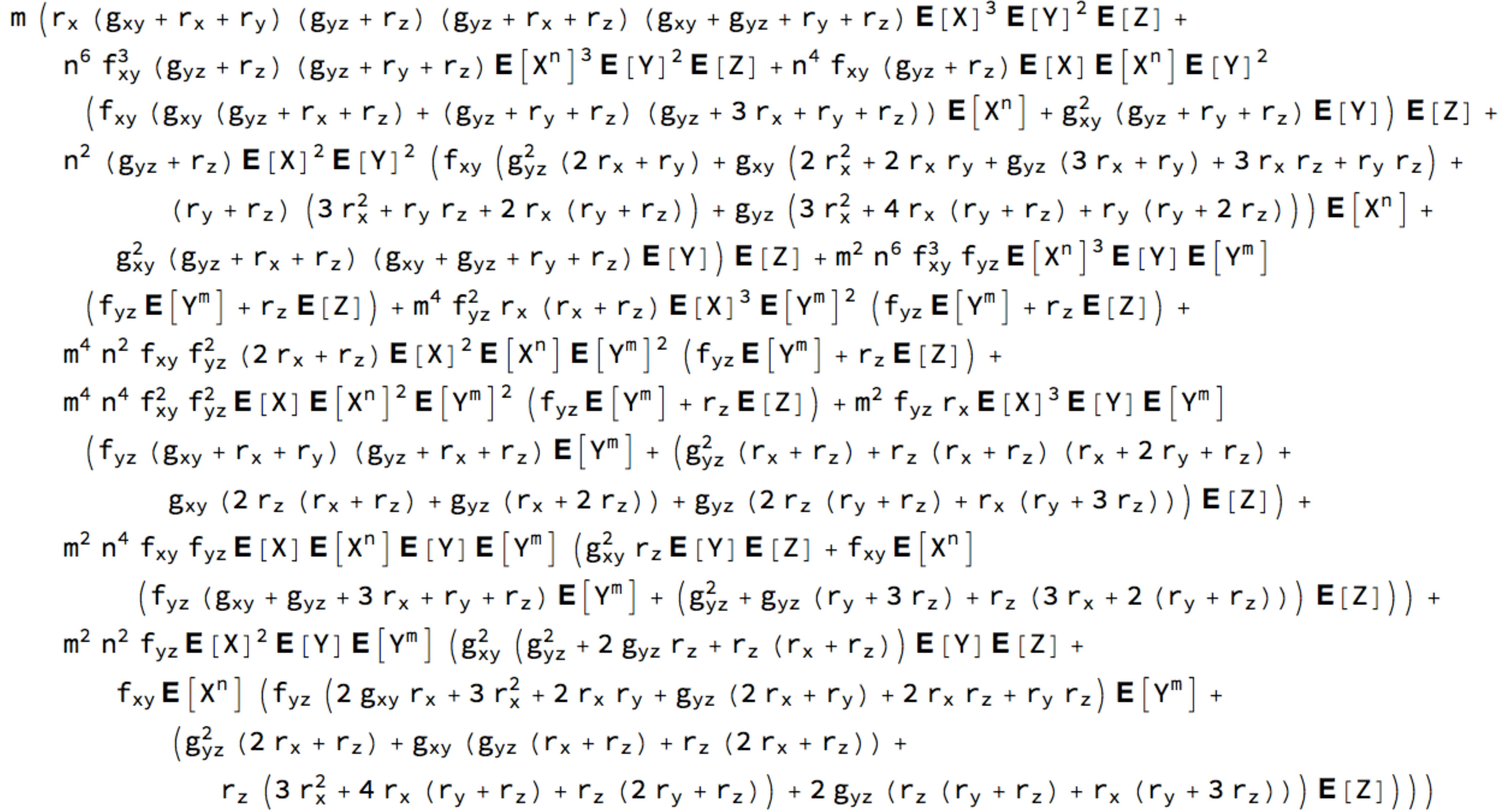}    
\label{gammayz}
\end{center}
\end{figure}

with $D=$
\begin{figure}[H]
\begin{center}
\includegraphics[width=13cm]{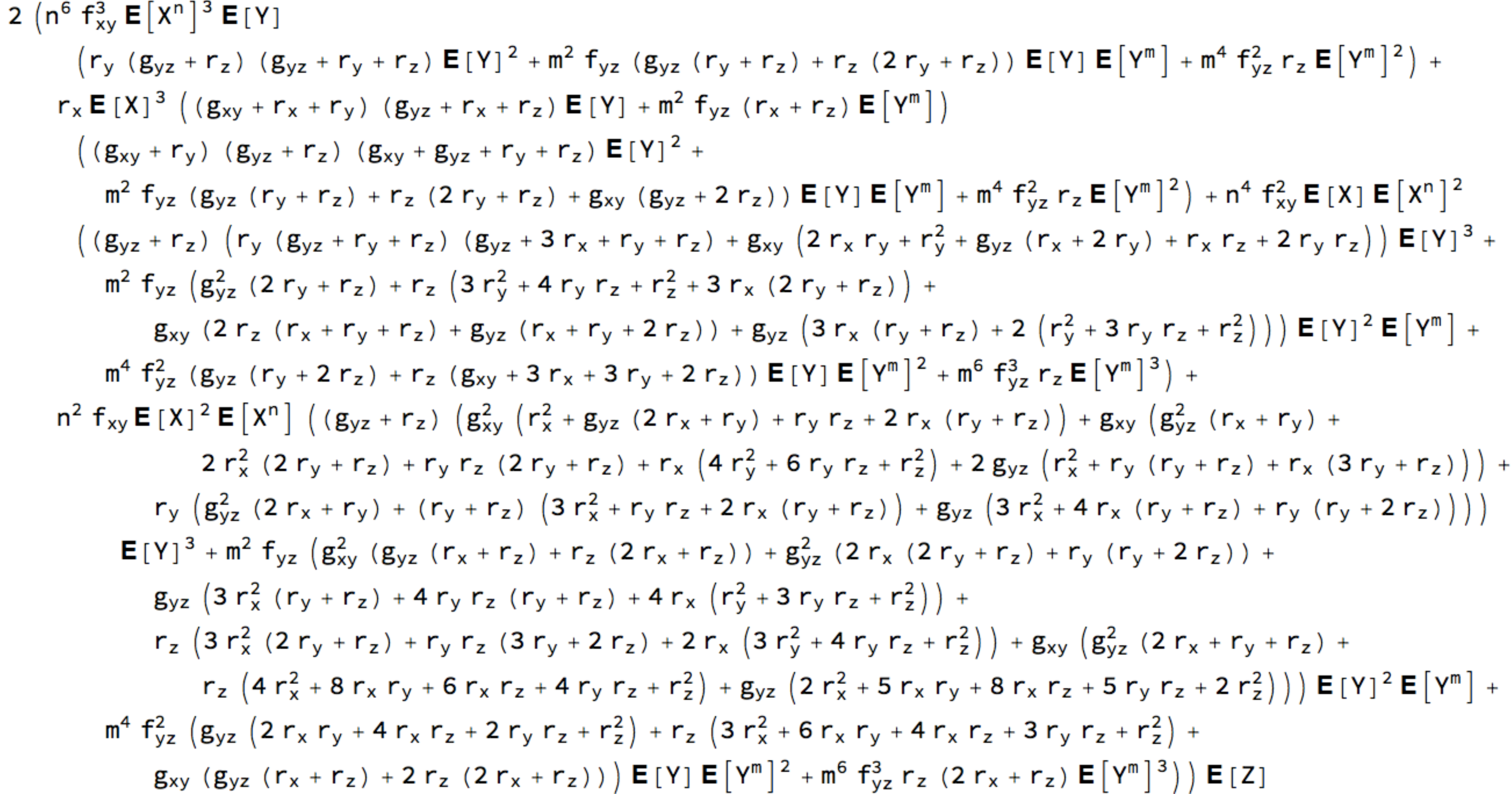}    
\label{gammaD}
\end{center}
\end{figure}

The $\gamma$ values that appear as coefficients in the individual Fano factors (Eqs (\ref{FFC1})) are obtained in a similar manner. They are not given here due to their complexity. 

\section*{Appendix B : Numerical Investigations}

The oligomerization reactions with an intermediate state of lower oligomeric order depend on a wide range of parameters, which makes them  complex to analyze even if the analytical solution is known. To limit the parameter space, 
we made the choice of considering only some parameters to be  free  and fixing the others. In particular, we analyzed in section 5 a CRN with fixed oligomerization degree, {\it i.e.} $n=2$ and $m=2$, which describes for example the tetramerization of monomeric proteins occurring through an intermediate dimerization step. 

For analyzing numerically the mean and  variances of the stochastic variables, we  also fixed all production rates $p_x=p_y=p_z$ to be equal to  $50$ and all degradation rates $r_x=r_y=r_z$ and $g_{xy}=g_{yz}$ to be equal to $0.001$. Other values have also been tested but  did not lead to  substantial differences in the interpretation of the results.  The free parameters considered were  $f_{xy}$ and $f_{yz}$ that describe the strength of the two dimerization reactions.  

\begin{figure}[h!]
\begin{center}
\includegraphics[width=9cm]{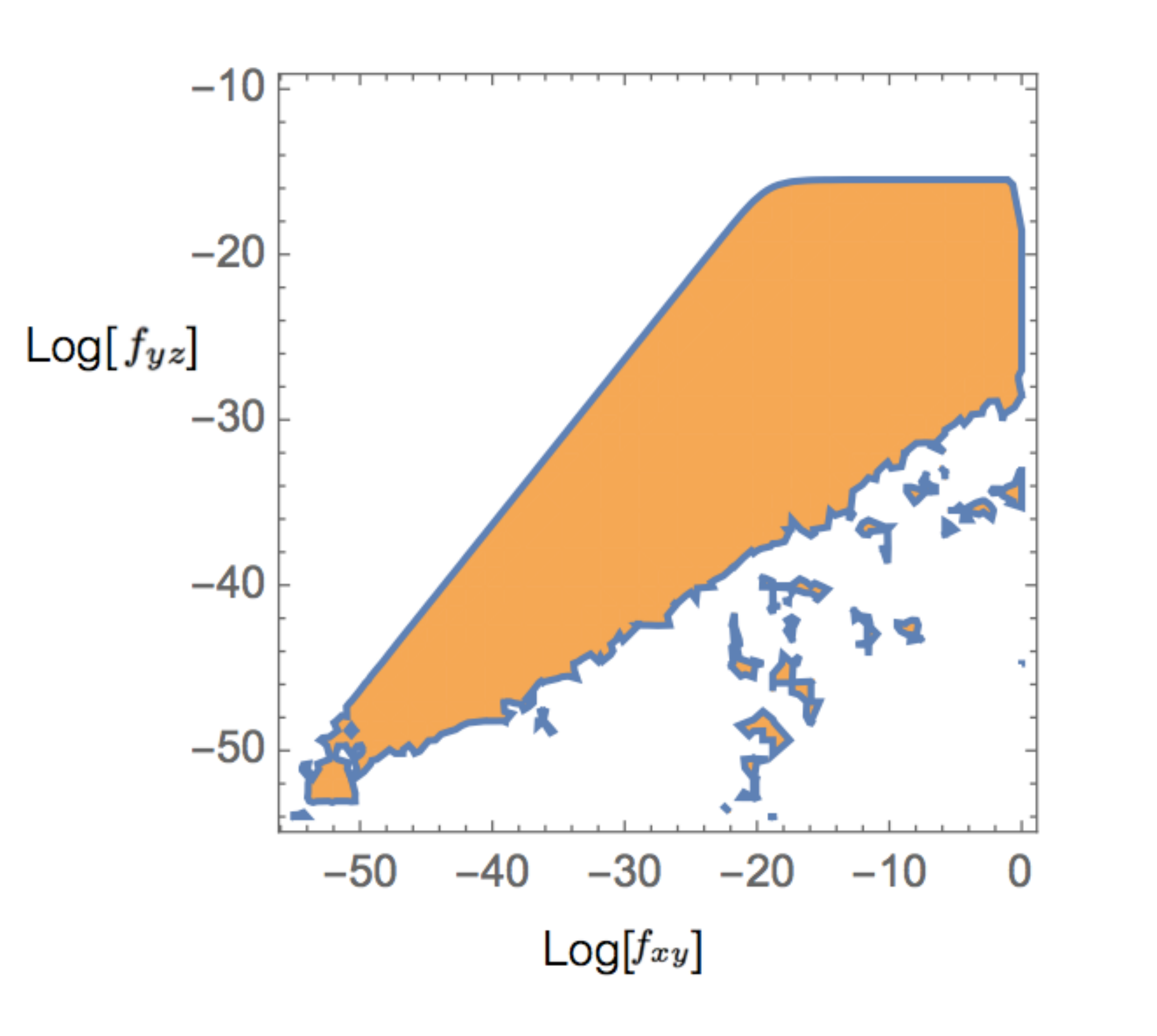}    
\caption{\bf{Domain of existence of the steady state solution} as a function of the logarithm of $f_{xy}$ and $f_{yz}$ for $p_x=p_y=p_z=50$ and $r_x=r_y=r_z=g_{xy}=g_{yz}=0.001.$}
\label{Appendix001}
\end{center}
\end{figure}

The first step consisted in analyzing the domain of existence of the analytic solutions at the steady state, where the mean numbers of molecules and variances  are positive for all molecular species ($\E(X)$, $\E(Y)$, $\E(Z)$, $\Var(X)$, $\Var(Y)$, $\Var(Z)>0$). The result of this analysis, namely the domain of existence of a solution in the $f_{xy}-f_{yz}$ plane, is plotted in Fig. (\ref{Appendix001}) using logarithmically rescaled  $f$-values.   

\begin{figure}[h!]
\begin{center}
\subfigure[]{\includegraphics[width=12.5cm]{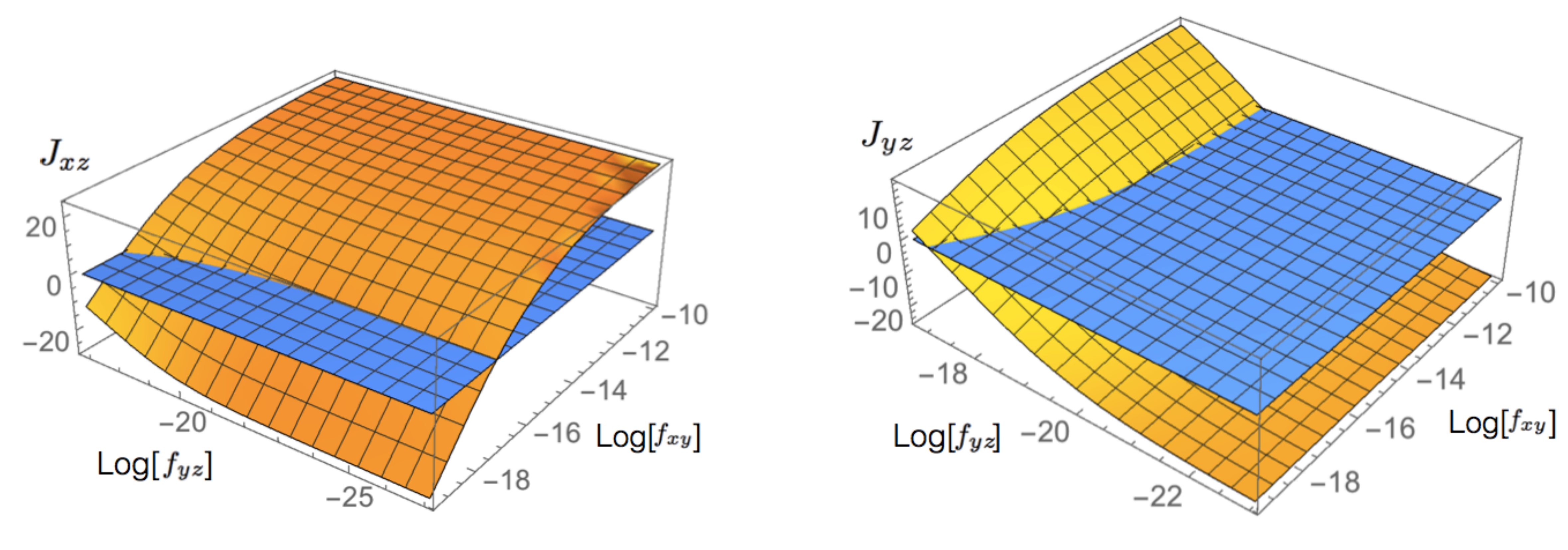}}  
\subfigure[]{\includegraphics[width=6cm]{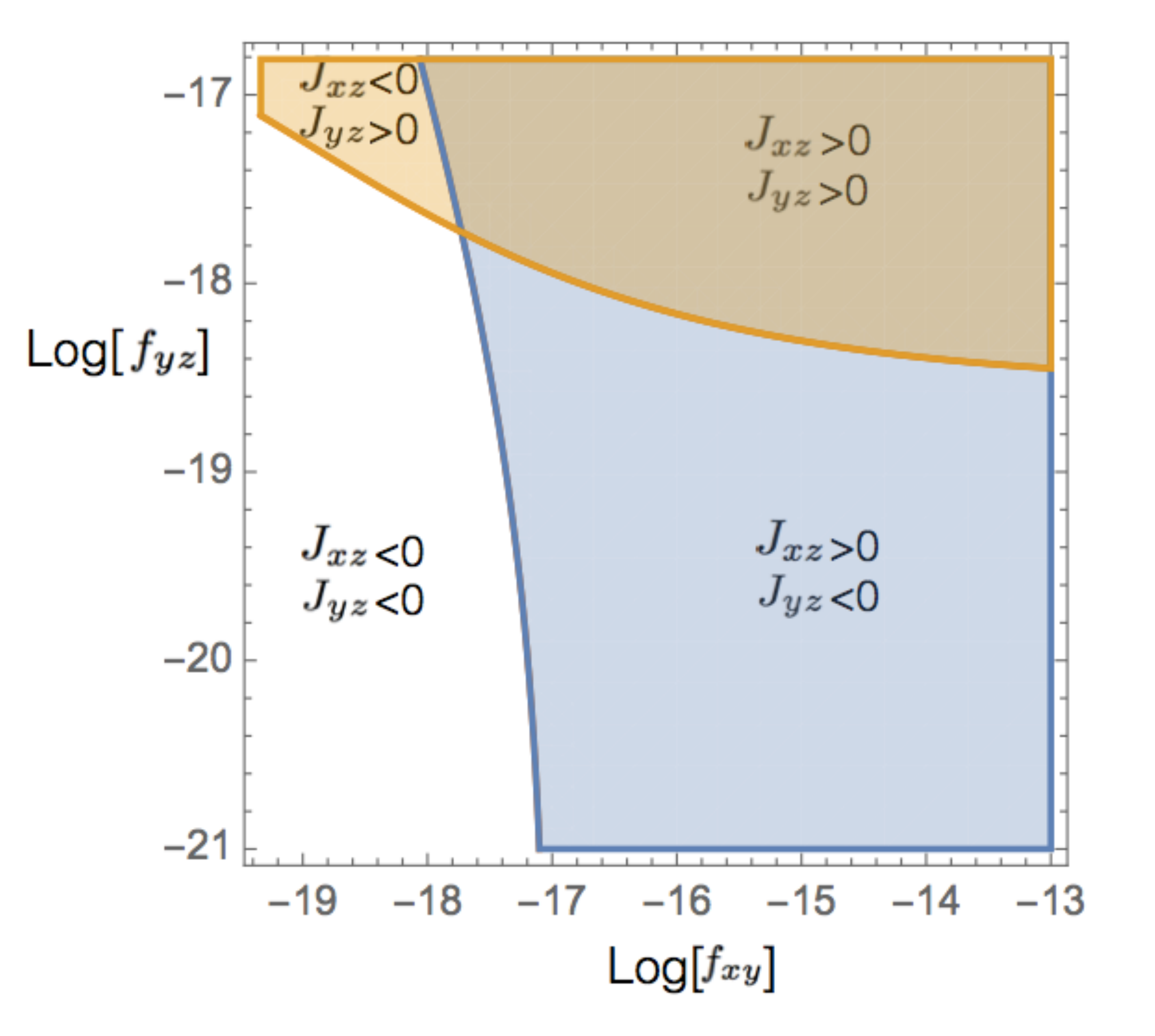}}    
\caption{{\bf Sign of the S-fluxes $J_{xy}$ and $J_{yz}$.} (a) $J_{xy}$ and $J_{yz}$ as a function of the free parameters $f_{xy}$ and $f_{yz}$; the blue surface is the plane that satisfies the equation $z=0$, and the orange planes are the two S-fluxes. (b) Projection of the two surfaces onto the $z=0$ plane. This projection divides the plane in four areas according to the signs of the S-fluxes. }
\label{appendix002}
\end{center}
\end{figure}

The second step consisted in analyzing the sign of the two independent S-fluxes, $J_{xy}$ and $J_{yz}$, in this domain. The results are shown in  Fig. \ref{appendix002}a. 

\noindent
By projecting the S-fluxes $J_{xy}(f_{xy},f_{yz})$ and $J_{xy}(f_{xy},f_{yz})$ onto the plane $z=0$, we can see for which parameter values  the sign of the fluxes are equal or differ  (Fig. \ref{appendix002}b). The four areas delimited in this way are referenced  in the main text and in Fig. \ref{Intermediate} as the four quadrants using Roman numerals.

\end{document}